\documentstyle[prl,aps,preprint]{revtex}
\begin{document}
\title{Metal-insulator transition in  VO$_{2}$: a 
Peierls-Mott-Hubbard mechanism} 

\author{Xiangyang Huang, Weidong Yang, and Ulrich Eckern} 
\address{Institut f\"ur Physik, Universit\"at Augsburg, 
D-86135 Augsburg, Germany}

\date{\today}
\maketitle
\begin{abstract}
The electronic structure of VO$_2$ is studied in the
frameworks of local density approximation (LDA) 
and LDA+$U$ to give a quantitative description of 
the metal-insulator (MI) transition in this system.
It is found that, both structural distortion and the 
local Coulomb interaction, play important roles in 
the transition. An optical gap, comparable to the 
experimental value has been obtained in the 
monoclinic structure by using the LDA+$U$ method. 
Based on our results, we believe that both, the 
Peierls and the Mott-Hubbard mechanism, are essential
for a description of the MI transition in this system.
 
\end{abstract}



The observation of a metal-insulator (MI) transition 
in VO$_2$ has attracted much interest for several years. 
Actually, this material has been considered a prototype 
of transition-metal oxides with an interesting MI 
transition. The transition occurs at temperature about 
340 K with a cell-doubling phase transition from the 
rutile to the monoclinic structure, its conductivity 
jumping by more than five orders of magnitude\cite{FJ}.
The transition is associated with a small 
change in volume, but significant internal atomic rearrangement 
characterized as both a pairing and an off-axis displacement 
of alternate vanadium atoms. In its insulating phase,
VO$_2$ has no magnetic order. The discovery of this 
transition initiated several experimental \cite{SS,EG,VM,TU} 
and theoretical investigations\cite{JB,AZ,EC,CS,RM,VE},
trying to explain the explanation of the mechanism of 
this phenomenon.

It is commonly accepted that Peierls and Mott-Hubbard 
pictures are two main mechanisms for MI transitions in 
half-filled band systems. Since VO$_2$ is of a 
half-filled band type, and possesses both a structural 
transition as well as a significant Coulomb interaction 
($U$) in the $3d$ bands of vanadiums, the question arises 
which mechanism drives the transition.
This problem has been debated for nearly 
four decades, and continues to be a controversy.
An early attempt was given by Goodenough\cite{JB}, who,
due to the existence of V-V pairings in the monoclinic 
structure, argued the transition to be of Peierls type: 
in his point of view, the V-V pairing 
and the pronounced octahedral distortion are responsible 
for the insulating state, and the gap is solely determined by 
the crystalline distortion. 
Somewhat later, Zylbersztejn and Mott proposed
another model\cite{AZ}.  With the information 
obtained from the properties of V$_{1-x}$Cr$_{x}$O$_{2}$, 
they argued that the band gap should essentially be due
to the Hubbard mechanism. 
Nevertheless, both of the two models are based on qualitative 
analysis and have not offered us a satisfied
scenario of the MI transition.

Many theoretical efforts have been made to give a 
quantitative description of the MI transition in VO$_{2}$ from 
first-principles calculations. Earlier works include a 
semiempirical band study by Caruthers et al.\cite{EC} 
and a cluster calculation on V-O octahedrons by 
Sommers et al.\cite{CS}.
Most noticeable, in 1994, an {\it ab initio} molecular 
dynamics (MD) study on VO$_2$ performed by Wentzcovitch 
et al.\cite{RM} gave a good description to the structural 
performance of VO$_{2}$, in which V-V pair distances 
along the c-axis in the monoclinic phase are further reduced
from the experimental value. 
However, their calculated optical gap of -0.04 eV 
contradicts the observed value of 0.7 eV\cite{SS}. 
A recent study based on the augmented spherical wave (ASW) 
method within the atomic sphere approximation (ASA)
gave nearly the same band picture of electronic 
structure\cite{VE}.

Thus up-to-date results from LDA and other
calculations can not provide a satisfactory quantitative 
description to the MI transition in VO$_{2}$.
We suspect that this failure is mainly due to the neglect of 
the local Coulomb interaction, which may be significant 
in 3$d$ transition oxides. Since no full potential, 
all-electron calculations, and no quantitative study 
considering a finite Coulomb interaction 
$U$ have been reported so far, the driving mechanism is still
unclear, and a comprehensive study with taking into 
account both effects is in order. In this article, 
we propose to study the mechanism of the MI 
transition in VO$_2$ by using the LDA as well as 
the LDA+$U$ method\cite{VI,AI}. 
The latter has been proved to be an efficient and reliable 
tool to study systems with strong Coulomb interactions.
Our results show that both, lattice distortion, and 
Coulomb interaction play important roles in the transition.
It seems that correct and quantitative description can 
not be obtained without taking into account either of them. 

Our calculations are based on density functional theory within 
LDA and LDA+$U$. The LDA calculations are performed in the 
full-potential linearized augmented plane-wave method (FP-LAPW)
with local orbital extensions\cite{DJ}, and the full-potential 
linear muffin-tin orbital approach (FP-LMTO)\cite{SY},
while LDA+$U$ calculations\cite{AI} are carried out within the 
framework of the FP-LMTO approach\cite{SY}.
The non-overlapping muffin-tin spheres are adopted in both 
approaches. For the FP-LMTO approach, three $\kappa$'s basis 
sets, -0.1, -1.0, and -2.5 Ry, are chosen. The potential is 
expanded in spherical harmonics inside the sphere,
and Fourier transformed in the interstitial region. Therefore,
no shape approximations are made to the density or potential.
In both methods, the calculated results are checked for 
convergence with respect 
to the number of {\bf k} points and the plane wave cutoff 
energy. In the LDA+$U$ method, the screened $U$ parameter used 
are ranging from 1.2 eV to 5.0 eV, which have been suggested 
by different authors, and are reasonable in early transition oxides 
\cite{SS,TU,SO,AE}. The screened $J$ is estimated to be 0.12 $U$.


Our calculations have been carried out in the rutile structure 
of VO$_2$ (space group $P4_{2}/mnm$) as well as in the monoclinic 
M$_1$ structure (space group $P2_{1}/c$). The lattice parameters
for these two structures used in our work are taken from McWhan 
et al.\cite{DB} and Longo et al.\cite{JM}, respectively.
In contrast to previous band calculations, our calculations were
performed with full-potential methods. 

The densities of states (DOS) of the two structures, as
determined by the LDA calculation, are shown
in Fig.~\ref{fig1}. It can be seen from Fig.~\ref{fig1}(a)
that the DOS in the metallic phase
near the Fermi level is contributed mainly from
vanadium $3d$ states, which are split by the
octahedral field into two groups, a lower triply
degenerate state of $t_{2g}$, and an upper doubly
degenerate state of $e_g$. The degeneracy of 
$t_{2g}$ ($e_g$) is further eliminated due to 
the orthorhombic field. Two distinct changes are obvious
when comparing the two phases. One is an upward shift of 
so-called $\pi^*$ bands ($d_{xz}$ and $d_{yz}$) 
above the Fermi level, the other is a splitting of the 
$d_{||}$ band ($d_{xy}$) of about 1.7 eV. 
We argue that the former
change is mainly caused by the V-V pairing.
According to our calculation, the shifts of
$d_{xz}$ and $d_{yz}$ band are about 0.18 and 0.27 eV
with respect to the rutile phase, respectively.
The occupation number of the $\pi^*$ band is
reduced from 0.45 e/atom in the rutile phase, to 
0.15 e/atom in the monoclinic phase. These two effects 
greatly eliminate band crossings, and lead to a reduction 
of the DOS at the Fermi level (see Fig.~\ref{fig1}(b)). Our 
calculations are in qualitative agreement with 
Goodenough's model\cite{JB} and quantitatively confirm
the recent results by Eyert\cite{VE}. 
The results obtained from the {\it ab initio} MD gave
a smaller overlap of bands near the Fermi level than ours: 
in the MD calculation, the potential adopted is a pseudo 
one and the parameter used is theoretically optimized,
i.e. the c-axis V-V pairing has been reduced from the 
experimental value of 2.62 \AA~to 2.58 \AA.

Our calculations show, confirming a suggestion by 
Eyert\cite{VE}, that further reducing the V-V pairing 
to $\sim$ 2.50 \AA~even produces a non-overlap band 
structure; which
even though the value is unrealistic, it is an interesting
observation in itself. Our FP-LAPW calculations also 
show that artificial reduction of 
V-V paring have raised the total energy of the monoclinic 
phase comparing to that of the experimental lattice parameter.
Specifically, the total energy in the experimental parameter is
calculated to be of $\sim$ 6 mRy/molecule lower compared to
the case when the optimized value is used\cite{RM}. 
We also used the generalized gradient approximation (GGA) 
to calculate the band structure of VO$_2$, with essentially no
improvement. We conclude that the structural distortion 
plays an important role
in MI transition in VO$_2$, and it strongly alters the band
structure of the insulating phase near the Fermi level.
The upward shift of the $\pi^*$ and the separation of 
the upper $d_{||}$ band from the lower $d_{||}$ band 
caused by the V-V pairing leaves the lower $d_{||}$ band 
nearly filled. Even so, it can only
lead to a reduction of carriers (or a small gap) near the
Fermi level. The electronic properties of VO$_2$ in the 
insulating state, therefore, can not be described 
by a LDA band calculation. A significant improvement is 
obtained, as we demonstrate below, by supplementing the
LDA by a local Coulomb term $U$.

In order to get a better picture of the MI transition in 
VO$_2$, we go beyond the LDA and check 
the part of correlation played in this system. 
The band structure of VO$_2$ in the insulating 
phase has been computed by the LDA+$U$ method,
and is shown in Fig.~\ref{fig2}. [$U$=0 represents the result 
for LDA.] We see that the band structure of the monoclinic 
phase is indeed modified by $U$. The Coulomb 
interaction $U$ has two effects on the band structure, see,
for example, Fig.~\ref{fig2}(b): the first effect is to 
cause a dramatic shift of the $\pi^*$ band from the lower 
$d_{||}$ band for $\sim$0.7 eV, the other one is to cause a 
further separation of the $d_{||}$ bands. 
The former effect is responsible for opening a gap;
the latter one is not so important since the 
separation of the $d_{||}$ band by the structural 
distortion has been large enough, i.e., greater than the 
optical gap of 0.7 eV. Notice that this can
only be achieved on the basis of the structural distortion.
To illustrate this point of view, we have performed a 
LDA+$U$ calculation on the metallic structure of VO$_2$,
with the result that a gap is not opened by a value of 
$U$ up to 7 eV. In this case, the overlap of the $\pi^*$ 
and the $d_{||}$ band  has not been eliminated,
and the $d_{||}$ band near the Fermi level is still of
a half-filled type. Further increasing $U$ is not
reasonable in this system. This clearly 
indicates that the metallic phase is determined by the 
structure, and can be adequately understood by LDA theory.

By varying $U$ in the reasonable range from 0 eV to 5 eV in
the insulating phase of VO$_2$, we obtain gap values
from -0.17 eV to 0.96 eV. The band gap as a function 
of the Coulomb interaction $U$ is displayed in Fig.~\ref{fig3}.
[A negative gap represents an overlap of the $d_{||}$ and 
the $\pi^*$ bands near the Fermi level.] 
A gap is opened, and exhibits roughly a linear relationship 
with $U$ in the range of $U$ between 1.6 eV and 5 eV. 
Specially, for $U$=4 eV, our calculation gives a gap of 
0.67 eV, which is in good agreement with the experimental 
value of 0.7 eV. It is instructive and necessary to apply
the LDA+$U$ method to MoO$_2$ and check whether this method
are valid in our study. Although MoO$_2$ crystallizes in the 
same monoclinic structure as low temperature VO$_2$, it is still
metallic. The LDA+$U$ band calculations\cite{XY} shows that 
the lower $d_{||}$ band is completely occupied, and that
the Fermi level is located in the lower-middle part of the
$\pi^*$ band. This condition is quite different from 
low temperature VO$_2$. Since the overlap of bands near the Fermi
level is about 1 eV, even a rather large $U$ (up to 6.0 eV)
can not open a gap in this system. This explains well why 
MoO$_2$ is a still metal in spite of large $U$ values.

Figure \ref{fig4} shows the calculated DOS by LDA+$U$ compared
with the ultraviolet photoemission spectroscopy (UPS)
spectrum\cite{SS} in the insulating phase of VO$_2$.
The UPS spectrum clearly shows that the $2p$ and $3d$ band
are in the energy range of -2 to -9 eV, and 0 to -2 eV, 
respectively.
We find that there is an overall good agreement between 
experiment and theory, with the exception of an underestimation of
the $3d$ downshift by about 0.9 eV. Also, the calculated DOS 
fits well with the x-ray photoemission spectroscopy (XPS) and UPS 
spectra by Goering et al.\cite{EG} and Bermudez et al. \cite{VM} 
First, the calculated DOS drops to zero at the Fermi level, 
which well explains the structure of UPS spectrum near
the Fermi level. Second, the three peaks in the DOS curve centered
at about -3.5 eV, -5.2 eV, and -7 eV are in good agreement with 
the O $2p$ structure in the UPS spectra. 
We also compared our result with the O $1s$ x-ray absorption 
spectroscopy (XAS) spectra\cite{MA} which reflects
the unoccupied states. The three peaks of $\pi^*$, $d_{||}$ and
$\sigma^*$ can also be well explained by our DOS results.

Further discussion about 3$d$ transition oxides such us
TiO$_2$ and CrO$_2$ will help us to understand the MI 
transition in VO$_2$. It is interesting to note that
the above oxides in rutile structure exhibit different 
properties, i.e., TiO$_2$ is an insolator, VO$_2$ undergoes
a MI transition and CrO$_2$ is a semi-metal.  A simple model 
derived from the band picture can give a good explanation.
There are two MO$_2$ (M=Ti, V, Cr) units per cell in
each systems while the numbers of occupied valence 
electrons in the $d_{||}$ bands are quite different.
In TiO$_2$, since the O $2p$ orbitals are filled,
the zero number of $d$ electrons leave unoccupied $d$ bands 
above Fermi level and cause an insulating phase.
This is quite different from the case of VO$_2$. There are two
half-filled bands near the Fermi level which can not be split
by a rather large value of $U$. This situation can be changed
by cell-doubled distortion during the MI transition. In the
insulating phase there are four VO$_2$ molecular per cell which
nearly fills the lower $d_{||}$ bands.
According to this picture, one may expect CrO$_2$ to be an 
insulator since the two bands near the Fermi level are fully 
occupied by four electrons, and the Coulomb interaction 
will separate these bands
from other $d$ bands. However, the ferromagnetic mechanism 
 will account for the semi-metal property\cite{KS,II}.

In conclusion, the band structure of VO$_2$ is calculated with the
LDA and LDA+$U$ methods. Our results show that both, lattice distortion
and Coulomb interaction, play important roles in the MI transition in
VO$_2$. First, the distortion strongly alters the band structure of
the low temperature monoclinic phase relative to that of the high
temperature
rutile phase. It induces a small overlap of bands at the Fermi
level but still leaves the system to be metallic. Second, the Coulomb
interaction further splits the $\pi^*$ bands from Fermi level
via the Mott-Hubbard mechanism. This cooperative effect causes an optical
gap and thus the MI transition. With a reasonable value of $U$ 
$\sim$ 4 eV,
we obtain a gap of 0.67 eV, which is close to that observed
in experiment. A similar calculation is also performed in the rutile
phase. There, no gap is found for $U$ up to 7 eV. In conclusion, neither
the Peierls nor the Mott-Hubbard mechanism provides a comprehensive
picture of the transition. Our calculation indicates that it is rather
a combination of both, which gives a quantitative description of the 
metal-insulator transition in VO$_2$.


One of us (XH) acknowledges useful discussions with Prof. Y. Wang, as
well as financial support by the Alexander von Humboldt-Stiftung.
We are grateful to Dr. S. Y. Savrasov for providing his FP-LMTO code.
This work was supported by the Deutsche Forschungsgemeinschaft
(DFG-Forschergruppe HO 955/2). The calculations were partly done on 
the IBM SP2 at the LRZ in Munich.

\begin{figure}
\caption{Partial V 3$d$ densities of states (DOS) of VO$_2$ in 
(a) the high temperature rutile and (b) the low temperature
monoclinic structure, respectively, within the LDA method.}
\label{fig1}
\end{figure}

\begin{figure}
\caption{Band structure of monoclinic VO$_2$ from (a) LDA and (b)
LDA+$U$, with $U$=4.0 eV.}
\label{fig2}
\end{figure}

\begin{figure}
\caption{Band gap vs screened (local) Coulomb interaction.
A negative "gap" means that an overlap of the $d_{||}$ and
the $\pi^*$ bands.}
\label{fig3}
\end{figure}

\begin{figure}
\caption{Comparision of ultraviolet photoemission data [2]
with the calculated DOS of monoclinic (insulating) VO$_2$.}

\label{fig4}
\end{figure}

\begin{references}
\bibitem{FJ}
F. J. Morin, Phys. Rev. Lett. {\bf 3,} 34 (1959).
\bibitem{SS}
S. Shin, S. Suga, M. Taniguchi, M. Fujisawa, H. Kanzaki, A. Fujimori,
H. Daimon, Y. Ueda, K. Kosuge, and S. Kachi, Phys. Rev B. {\bf 41,} 
4993 (1990).
\bibitem{EG}
E. Goering, M. Schramme, O. M\"uller, R. Barth, H. Paulin, M. Klemm
M. L. denBoer and S. Horn, Phys. Rev. B {\bf 55,} 4225 (1997).
\bibitem{VM}
V. M. Bermudez, R. T. Williams, J. P. Long, R. K. Reed, 
and P. H. Klein, Phys. Rev. B {\bf 45,} 9266 (1992).
\bibitem{TU}
Takayuki Uozumi, Kozo Okada, and Akio Kotani, J. of Phys. Soc. Jpn.
{\bf 62,} 2595 (1993).
\bibitem{JB}
J. B. Goodenough, Phys. Rev. {\bf 117,} 1442 (1960).
\bibitem{AZ}
A. Zylbersztejn and N. F. Mott, Phys. Rev. {\bf B 11,}
4383 (1975). 
\bibitem{EC}
E. Caruthers, L. Kleinman, and H. I. Zhang, Phys. Rev. B 
{\bf 7,} 3753 (1973).
\bibitem{CS}
C. Sommers, R. de Groot, D. Kaplan, and A. Zylbersztejn,
J. Phys. Lett. {\bf 36,} L157 (1975).
\bibitem{RM}
R. M. Wentzcovitch, W. W. Schulz, and P. B. Allen, Phys. Rev. Lett.
{\bf 72,} 3389 (1994).
\bibitem{VE}
V. Eyert, to be published.
\bibitem{VI}
V. I. Anisimov, J. Zaanen, and O. K. Andersen, Phys. Rev. B
{\bf 44,} 943 (1991).
\bibitem{AI}
A. I. Liechtenstein, V. I. Anisimov, and J. Zaanen, Phys. Rev. B
{\bf 52,} R5467 (1995).
\bibitem{DJ}
D. J. Singh, {\it Planewaves, Pseudopotentials and the LAPW method}
(Kluwer Academic Publishers, Boston, 1994).
\bibitem{SY}
S. Y. Savrasov and D. Y. Savrasov, Phys. Rev. B {\bf 46,} 12181 (1992).
\bibitem{SO}
C. Sommers and S. Doniach, Solid State Commun. {\bf 28,} 133 (1978).
\bibitem{AE}
A. E. Bocquet, T. Mizokawa, K. Morikawa, A. Fujimori, 
S.R. Barman, K. Maiti, D. D. Sarma, Y. Tokura, and 
M. Onoda, Phys. Rev B {\bf 54,} 5368 (1996).
\bibitem{DB}
D. B. McWhan, M. Marezio, J. P. Remeika, and P. D. Dernier, 
Phys. Rev. B {\bf 10,} 490 (1974).
\bibitem{JM}
J. M. Longo, and P. Kierkegaard, Acta Chim. Scand. {\bf 24,}
420 (1970).
\bibitem{XY}
X. Y. Huang (unpublished).
\bibitem{MA}
M. Abbate, F. M. F. de Groot, J. C. Fuggle, Y. J. Ma, C. T. Chen,
F. Sette, A. Fujimori, Y. Ueda, and K. Kosuge,
Phys. Rev. B {\bf 43,} 7263 (1991).
\bibitem{KS}
K. Schwarz, J. Phys. F {\bf 16,} L211 (1986).
\bibitem{II}
I. I. Mazin, D. J. Singh, and C. Ambrosch-Draxl, cond-mat/9806378.

\end{references}
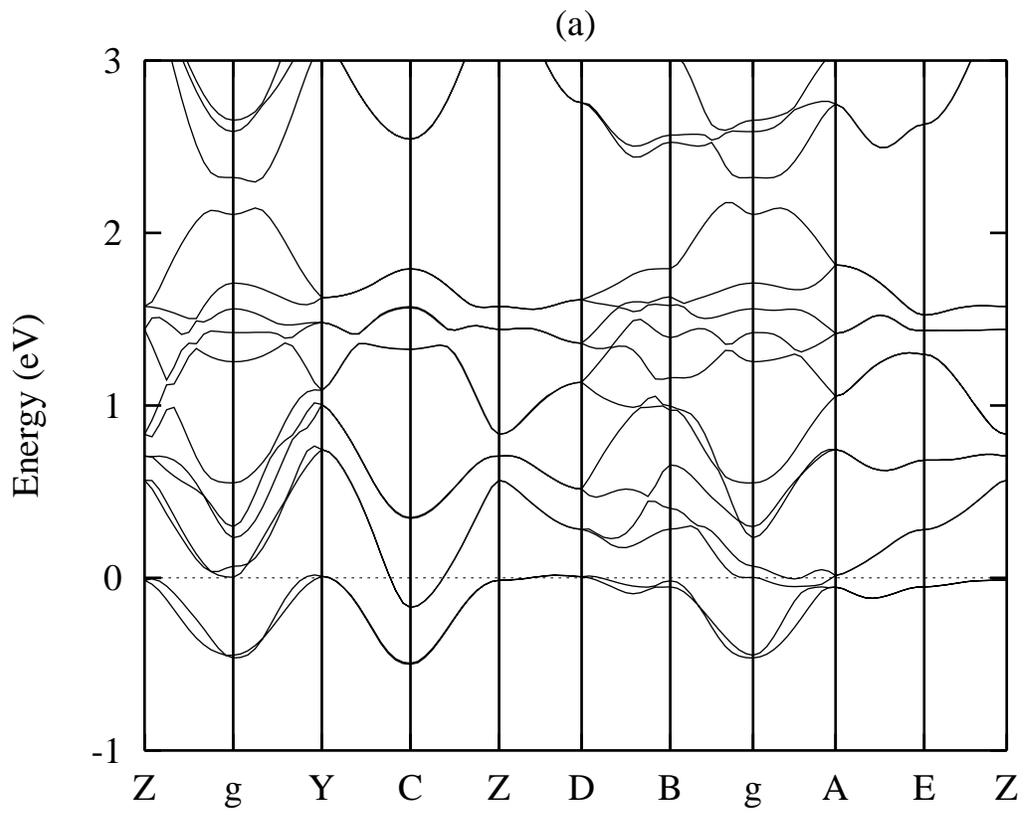
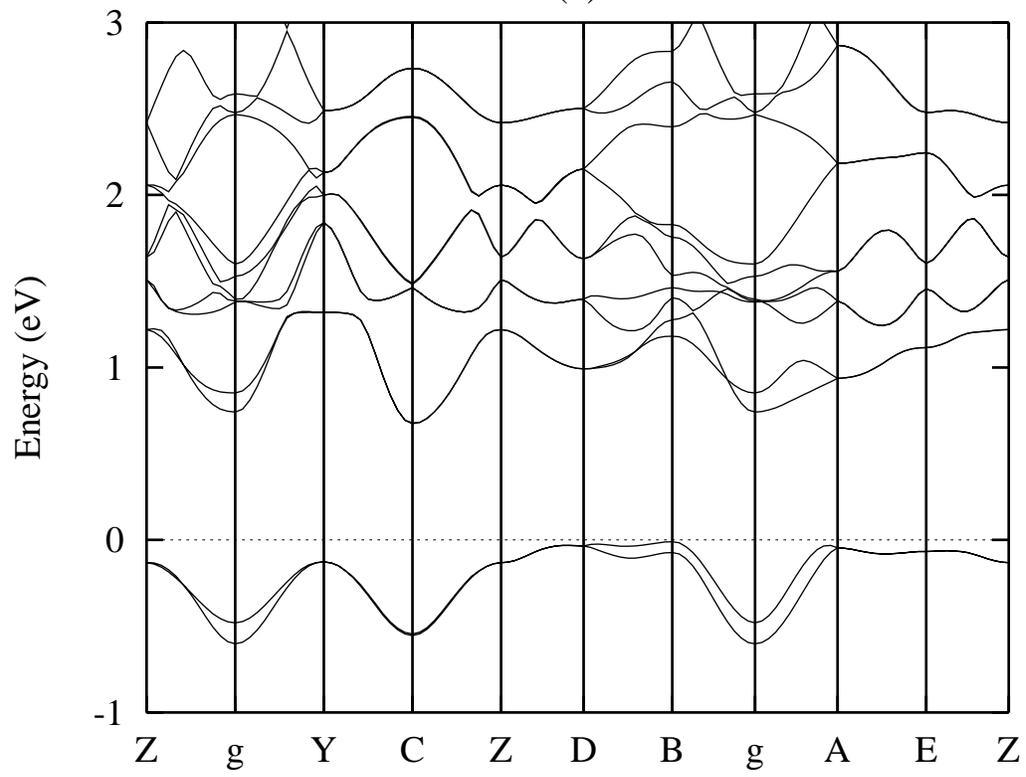
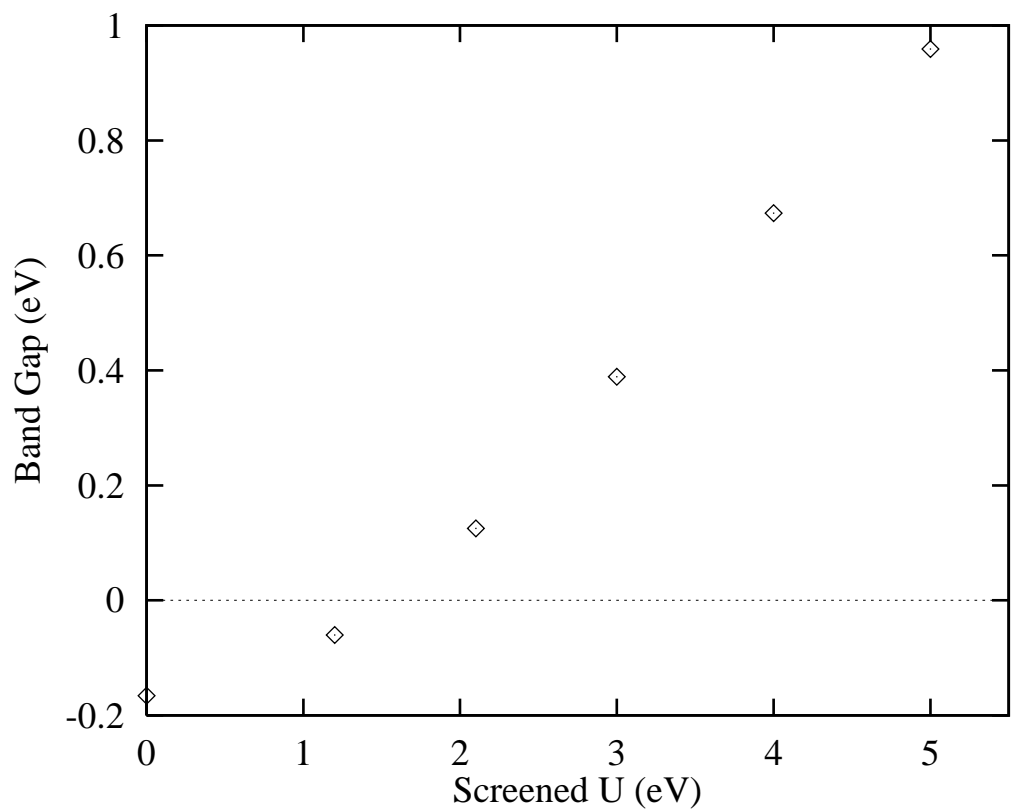
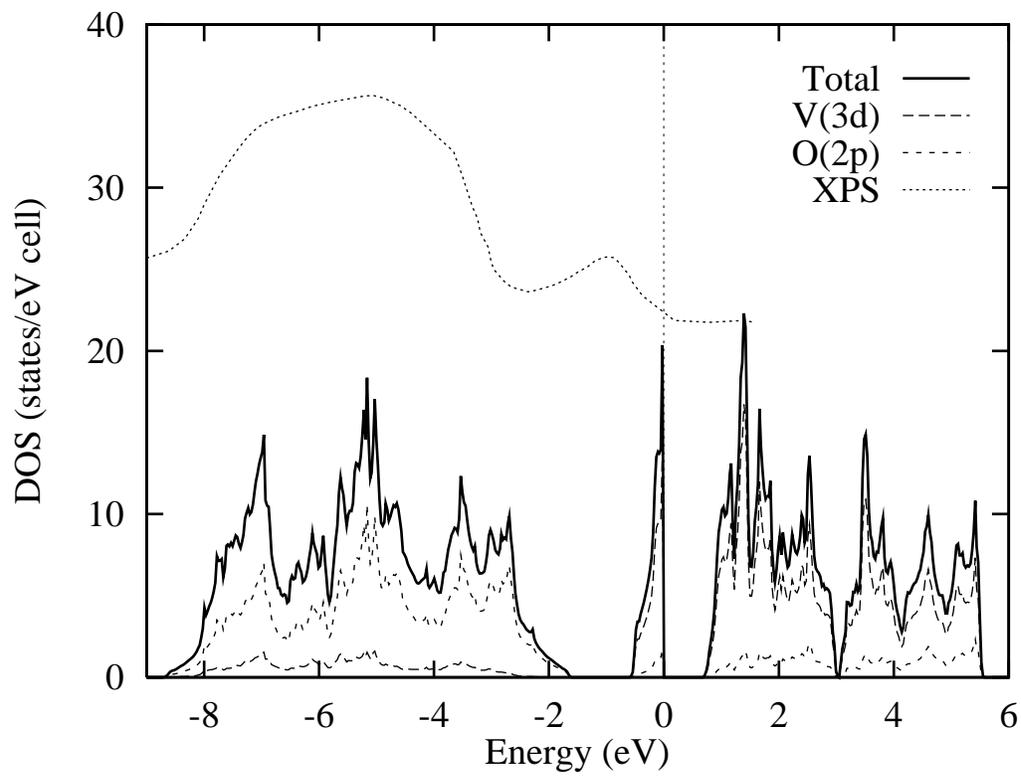
\end{document}